\documentclass[lettersize,journal]{IEEEtran}
\usepackage{amsmath,amsfonts}
\usepackage[noend]{algpseudocode}
\usepackage{array}
\usepackage{textcomp}
\usepackage{stfloats}
\usepackage{url}
\usepackage[hidelinks,colorlinks=true,linkcolor=blue,citecolor=blue,urlcolor=black]{hyperref}
\usepackage{acronym}
\usepackage{url}
\usepackage{tabularray}
\usepackage{verbatim}
\usepackage{subfigure}
\usepackage{amssymb}
\usepackage{graphicx}
\usepackage[dvipsnames]{xcolor}
\usepackage{cite}
\usepackage{xfrac}
\usepackage{multirow}
\usepackage{multicol}
\usepackage{matlab-prettifier}
\usepackage{mathtools}

\input{AcronymList}
\usepackage[linesnumbered,ruled,vlined]{algorithm2e}

\SetCommentSty{mycommfont}

\begin{document}
\title{Phase-Shifted Pilot Design for NOMA-Empowered Uplink ISAC Systems}

\author{Ahmet Sacid S\"{u}mer, Ebubekir Memi\c{s}o\u{g}lu, and H\"{u}seyin Arslan,~\IEEEmembership{Fellow,~IEEE}
\thanks{

Authors are with the Department of Electrical and Electronics Engineering, Istanbul Medipol University, Istanbul, 34810, Turkey (e-mail: ahmet.sumer@std.medipol.edu.tr; huseyinarslan@medipol.edu.tr).

Ebubekir Memişoğlu is 
also with the Department of Electrical and Electronics Engineering, Samsun University, Samsun, 55060, Turkey (e-mail: ebubekir.memisoglu@samsun.edu.tr). \\
}
}

\maketitle

\begin{abstract}
The deployment of multiple transmitters (TXs) in integrated sensing and communication (ISAC) networks necessitates efficient resource sharing to overcome the limitations of orthogonal allocation. While conventional interleaved (CI) pilots combined with non-orthogonal multiple access (NOMA) improve spectral efficiency (SE), they inherently compromise sensing resolution due to spectral sparsity, rendering the CI nulling (CIN) extension a strictly limited remedy. This paper proposes a phase-shifted (PS) pilot design and its novel PS nulling (PSN) variant to integrate a communication TX (CTX) over the PS-ISAC framework. The PSN variant strategically punctures sensing signals at CTX pilot locations to preserve initial channel estimates, enabling a dense data overlay. To resolve the resulting multi-TX interference, joint iterative interference cancellation (IIC) is adapted for non-nulling configurations and sequential IIC is adapted for nulling variants, optimizing for both detection robustness and convergence speed. Simulation results across varying STX densities and modulation orders demonstrate that the phase-shifted frameworks maintain sensing integrity while explicitly reducing receiver-side computational complexities by $18.8\%$ and $21.0\%$ against their respective interleaved baselines. 
\end{abstract}


\section{Introduction}
\label{sec:introduction}
\IEEEPARstart{T}{he} architectural evolution of \ac{6G} wireless networks is fundamentally predicated on the seamless convergence of communication and environmental sensing capabilities~\cite{liu2022evolution, yazar20206g}. \Ac{ISAC} has emerged as a pivotal paradigm to enhance \ac{SE}, minimize hardware redundancy, and reduce signaling latency by unifying these dual functionalities within a single transceiver architecture~\cite{wei2022toward, cui2021integrating}. This integration is highly anticipated for emerging applications requiring robust situational awareness and massive connectivity across complex topologies, fundamentally encompassing autonomous \ac{V2X} networks, cooperative relay-assisted architectures~\cite{IoT6_Coop_ISAC}, heterogeneous \ac{IoT} environments augmented by active \acp{RIS}~\cite{IoT3_RIS_Heto}, and high-mobility \ac{UAV} deployments~\cite{IoT4_OTFS}. Recently, \ac{CSI}-based methods have proven highly effective for achieving sensing goals within existing wireless networks, successfully demonstrating capabilities such as device-free human activity recognition~\cite{wang2017device} and vehicle speed estimation~\cite{wang2018device}.

However, the ubiquity of wireless sensing introduces a practical network challenges for dense \ac{IoT} systems. When sensing signals are utilized to extract \ac{CSI} in active networks, they inevitably collide with communication signals, creating severe mutual interference~\cite{10736552, zhang2022practical, zheng2019radar}. To resolve this interference, orthogonal resource allocation techniques utilizing time, frequency, or spatial division have been proposed, such as orthogonal spectral sharing algorithms~\cite{qian2022robust, qian2022radar}. Nevertheless, these orthogonal schemes are structurally inefficient at mitigating congestion and often degrade either the transmission rate or sensing performance of resource-constrained nodes. Consequently, the design of efficient multiple access schemes that can accommodate massive multi-\ac{TX} connectivity without unnecessarily degrading sensing performance or communication throughput has become an important research frontier for \ac{IoT} ISAC deployments~\cite{mu2022noma}.

To transcend these orthogonal limitations, recent literature has explored advanced methodologies, including dual-domain waveform superposition~\cite{tagliaferri2023integrated}, joint pilot and transmission optimization~\cite{hua2024integrated}, and mutual information-driven pilot designs~\cite{bazzi2025mutual, keskin2025fundamental, zhang2021overview}. Consequently, within the framework of next-generation multiple access~\cite{liu2024next}, the research community has increasingly pivoted toward non-orthogonal solutions to fundamentally maximize resource utilization.

\subsection{Prior Work and Motivation}
To directly address the spectral scarcity bottleneck, \ac{NOMA} has emerged as a critical enabler by permitting multiple \acp{TX}, encompassing both \ac{CTX} and \acp{STX}, to simultaneously share identical time-frequency resources~\cite{ahmed2024unveiling, makki2020survey}. Various \ac{ISAC}-\ac{NOMA} frameworks have been developed to manage the mutual interference, explicitly incorporating power-domain successive interference cancellation for full-duplex transceiver architectures~\cite{wang2022noma, IoT5_FD_NOMA}, the non-orthogonal coexistence of \ac{OFDM} and \ac{FMCW} signals~\cite{csahin2020multi}, and advanced multi-domain \ac{OTFS} modulation schemes~\cite{cui2022multi, IoT4_OTFS}. Furthermore, to support massive connectivity, recent literature has successfully expanded these non-orthogonal principles into near-field extremely large-scale multiple-input multiple-output frameworks, leveraging spatial degrees of freedom to dynamically cluster \ac{IoT} devices~\cite{IoT1_XL_MIMO}.

Advancing beyond these heterogeneous waveform strategies, foundational \ac{CSI}-based \ac{ISAC}-\ac{NOMA} architectures established the concept of non-orthogonally superimposing unified \ac{OFDM} signals for an isolated \ac{CTX} and \ac{STX} pair. By utilizing an \ac{IIC} \ac{RX}, this architecture successfully disentangled the mutual interference, achieving \ac{SE} perfectly comparable to interference-free benchmarks~\cite{memisoglu2023csi}.

Expanding these architectures to support multiple \acp{STX} is achieved via the \ac{CI-ISAC} pilot allocation strategy~\cite{demir2023csi}. However, \ac{CI-ISAC} inherently degrades the maximum unambiguous range due to spectral sparsity~\cite{10561589, mura2024optimized}. Furthermore, interleaved allocations demand a separate \ac{IFFT} for every scheduled \ac{STX} at the \ac{RX} side, creating a substantial computational burden as the network scales~\cite{APS-ISAC, sohl2007comparison}. Although a spectral nulling technique has been proposed for single-\ac{STX} scenarios to protect the communication payload by explicitly puncturing sensing pilots~\cite{zheng2025csi}, extending this mechanism to support multiple \ac{STX} nodes yields the \ac{CIN} baseline. While this multi-\ac{STX} extension is structurally straightforward, it inherits the fundamental structural trade-offs of interleaved designs.

To overcome these limitations, the \ac{PS-ISAC} pilot design~\cite{PS-ISAC} and its adaptive extension, \ac{APS-ISAC}~\cite{APS-ISAC}, utilize full-bandwidth pilot transmissions modulated by \ac{STX}-specific cyclic phase shifts. These structures ensure that multiple sensing \acp{CIR} remain perfectly separable in the delay domain while significantly reducing computational complexity. Specifically, the \ac{PS-ISAC} architecture requires only a single composite \ac{IFFT} at the \ac{RX} regardless of the number of scheduled \acp{STX}, a saving that is multiplied across iterations~\cite{APS-ISAC}. Despite maximizing range resolution and unambiguous range, phase-shifted designs have previously been restricted to orthogonal contexts. This paper bridges this gap by integrating the \ac{PS-ISAC} structure into a \ac{NOMA}-empowered system, adapting joint and sequential \ac{IIC} \acp{RX} to resolve complex multi-\ac{TX} interference within dense environments.

\subsection{Contributions}
To address the fundamental sensing-communication trade-offs in multi-\ac{TX} environments, the PS-\ac{ISAC}-\ac{NOMA} framework and its spectral nulling variant, \ac{PSN}-\ac{ISAC}-\ac{NOMA}, are proposed. The main contributions of this work are summarized as follows:

\begin{itemize}
\item 
We introduce an \ac{UL} architecture that overlays high-rate communication data onto full-bandwidth phase-shifted sensing pilots. By assigning \ac{STX}-specific cyclic phase rotations based on the \ac{CP} duration, the framework maps each sensing \ac{CIR} into a distinct and non-overlapping delay-domain window. 
We further introduce the PSN variant, which utilizes spectral nulling at communication pilot locations to protect the integrity of initial channel estimates.

\item 
We adapt and evaluate two distinct \ac{IIC} architectures to resolve the dense multi-\ac{TX} interference. A joint \ac{IIC} is utilized as a robust solution for non-nulling configurations where standard joint detection fails to resolve the multi-\ac{STX} interference. Additionally, we adapt a sequential \ac{IIC} specifically for the spectral nulling variants. 

\item 
We conduct an empirical comparison between the proposed phase-shifted frameworks and established CI and \ac{CIN} baselines across multiple modulation orders and \ac{STX} densities. Performance is assessed in terms of sensing \ac{NMSE}, communication \ac{NMSE}, \ac{BER}, \ac{SE}, and computational complexity. By enabling multi-\ac{STX} sensing through time-domain phase shifts instead of frequency-domain interleaving, the proposed approach substantially reduces \ac{RX}-side hardware complexity requirements.
\end{itemize}

\subsection{Notations}
Bold uppercase and lowercase letters denote matrices and vectors, respectively (e.g., $\mathbf{X}$ and $\mathbf{x}$). Non-bold letters represent scalar values, and $\mathbf{0}$ denotes an all-zero vector. Calligraphic symbols, such as $\mathcal{K}$ and $\mathcal{M}$, denote sets. The Dirac delta function is denoted by $\delta(\cdot)$. The constellation slicing operator $\mathcal{Q}(\cdot)$ maps soft estimates to the nearest points within the modulation alphabet $\mathcal{A}$. The superscripts $(\cdot)^H$ and $(\cdot)^T$ denote the Hermitian transpose and transpose operations, respectively. As a general convention, the notation $\hat{(\cdot)}$ represents refined estimated variables at the \ac{RX}. The notations $\tilde{(\cdot)}$ and $\breve{(\cdot)}$ denote intermediate interference-cancelled signals and composite multi-\ac{TX} variables prior to domain-specific separation, respectively. The absolute value is denoted by $|\cdot|$, while $\otimes$ and $\oslash$ denote element-wise Hadamard multiplication and division, respectively. The notation $\mathcal{CN}(\mu, \sigma^2)$ represents a circularly symmetric complex Gaussian random variable with mean $\mu$ and variance $\sigma^2$. To ensure taxonomic precision, the suffix -\ac{ISAC} designates sensing-only configurations, whereas -\ac{ISAC}-\ac{NOMA} denotes the non-orthogonal superposition of \ac{CTX} and \ac{STX} sequences. Baseline acronyms used in isolation, specifically PS, PSN, CI, and CIN, inherently imply the fully integrated -\ac{ISAC}-\ac{NOMA} framework.

\begin{figure}
\centering
\includegraphics[width=1\linewidth]{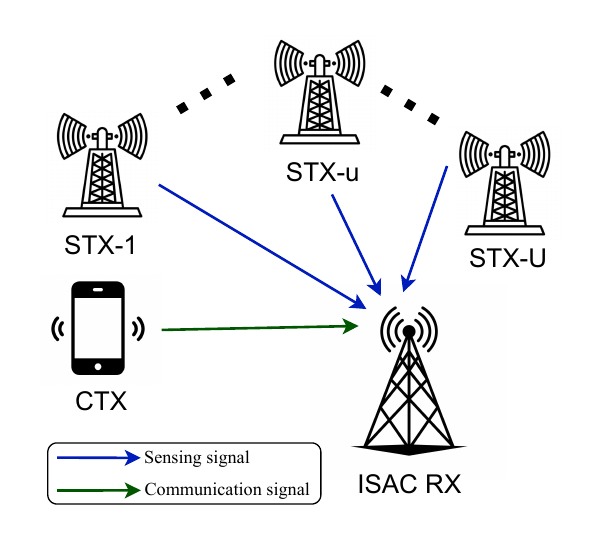}
\caption{System model of the proposed \ac{UL} \ac{ISAC}-\ac{NOMA} scheme.}
\label{fig:system_model}
\vspace{-3mm}
\end{figure}

\section{System Model}
\label{sec:system_model}

As illustrated in Fig. \ref{fig:system_model}, an \ac{UL} \ac{ISAC}-\ac{NOMA} scenario is considered. The architecture comprises a single-antenna \ac{CTX} and $U$ single-antenna \acp{STX}. All \acp{TX} simultaneously send \ac{OFDM} signals to the ISAC \ac{RX} over shared time-frequency resources. Consequently, the ISAC \ac{RX} receives the non-orthogonally superimposed signals to perform \ac{UL} channel estimation and data detection.

\subsection{Transmitted Signal Model}
The system operates over a frame of $M_{\text{sym}}$ \ac{OFDM} symbols utilizing $N$ subcarriers. Let $\mathcal{K} = \{0, \dots, N-1\}$ denote the set of subcarrier indices. To facilitate non-orthogonal coexistence across the shared subcarrier grid, the transmission architecture formally partitions the generated signals into a dedicated communication payload and their respective sensing waveforms.

\subsubsection{Communication Transmitter (CTX)}
The \ac{CTX} transmits a payload denoted by the frequency-domain vector $\mathbf{x}_{F,c} \in \mathbb{C}^{N \times 1}$. This payload consists of a mixture of information-bearing data tones and known pilot tones distributed across the subcarrier grid. Let $\mathcal{K}_p \subset \mathcal{K}$ and $\mathcal{K}_d \subset \mathcal{K}$ denote the disjoint sets of subcarrier indices allocated for the communication pilots and data symbols, respectively, such that $\mathcal{K}_p \cup \mathcal{K}_d = \mathcal{K}$. 

\begin{figure}
\centering
\subfigure[CI-ISAC-NOMA.]{\includegraphics[width=0.99\columnwidth , trim={1.8cm 0.25cm 1cm 0.25cm}, clip]{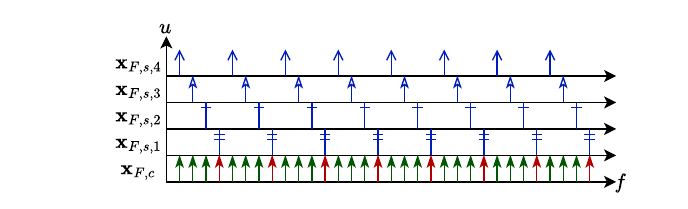}}
\subfigure[CIN-ISAC-NOMA.]{\includegraphics[width=0.99\columnwidth, trim={1.8cm 0.25cm 1cm 0.25cm}, clip]{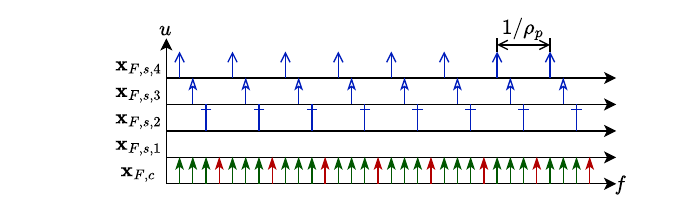}}
\subfigure[PS-ISAC-NOMA.]{\includegraphics[width=0.99\columnwidth, trim={1.8cm 0.25cm 1cm 0.25cm}, clip]{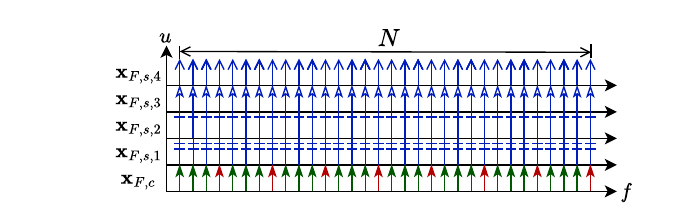}}
\subfigure[PSN-ISAC-NOMA.]{\includegraphics[width=0.99\columnwidth, trim={1.8cm 0.25cm 1cm 0.25cm}, clip]{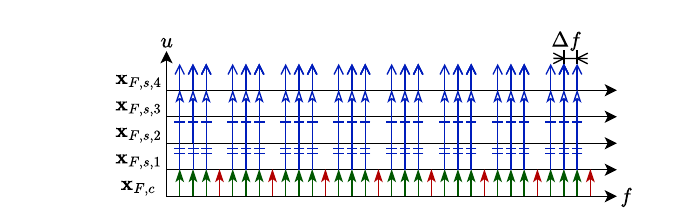}}
\caption{Frequency-domain resource allocation for \ac{UL} \ac{ISAC}-\ac{NOMA} with communication signal as the base layer. Sensing pilots are overlaid via: (a) CI: Interleaved with communication pilots; (b) CIN: Interleaved with nulling at communication pilot locations; (c) PS: Superimposed across the frame using cyclic phase shifts; and (d) \ac{PSN}: Phase-shifted with spectral nulling at communication pilot locations.}
\label{fig:Tx_Signals_Separate}
\vspace{-3mm}
\end{figure}

\subsubsection{Sensing Transmitters (STXs)}
To support multi-\ac{TX} sensing, four distinct pilot structures are evaluated, as illustrated in Fig.~\ref{fig:Tx_Signals_Separate}. The CI-ISAC-NOMA and CIN-ISAC-NOMA schemes allocate disjoint, sparse subsets of subcarriers to each \ac{STX}. While these methods provide an effective foundation for multi-\ac{STX} integration, their sparse nature inherently limits the maximum unambiguous range. 
In contrast, the proposed PS-ISAC-NOMA framework utilizes the full available bandwidth for each \ac{STX}. Building upon this foundational structure, the PSN-ISAC-NOMA architecture is introduced as its dedicated spectral nulling variant. This variant intentionally punctures the sensing sequence at the communication pilot subcarriers to preserve the initial \ac{CTX} channel estimates without sacrificing the underlying delay-domain separability. Let $\mathbf{x}_{F,r} \in \mathbb{C}^{N \times 1}$ denote a common reference probing sequence. To ensure separability at the ISAC \ac{RX}, a \ac{STX}-specific cyclic phase shift is applied. The frequency-domain sensing vector for the $u$-th \ac{STX}, denoted as $\mathbf{x}_{F,s,u}$, is defined by applying a diagonal phase rotation matrix $\mathbf{P}_u(\theta_u) \in \mathbb{C}^{N \times N}$, as follows: 
\begin{align}
    \mathbf{x}_{F,s,u} &= \mathbf{P}_u(\theta_u) \mathbf{x}_{F,r}, \nonumber \\
    \mathbf{P}_u(\theta_u) &= \text{diag}\left(e^{j\theta_{u,0}}, e^{j\theta_{u,1}}, \dots, e^{j\theta_{u,N-1}}\right), \label{eq:phase_shift_matrix}
\end{align}
where the subcarrier-dependent phase shift $\theta_{u,k}$ for the $k$-th subcarrier is defined as
\begin{align}
    \theta_{u,k} &= -2\pi k \frac{n_u}{N}. \label{eq:phase_shift_angle}
\end{align}
The effective time-domain shift $n_u$ is structurally parameterized by the \ac{CP} duration $N_{cp}$. Since $N_{cp}$ is inherently known at both the transmit and receive nodes, no additional complex signaling is required. The only requirement is conveying the phase-shift index to each scheduled \ac{STX}, similar to the assignment of interleaved subcarrier offsets in existing standards~\cite{3gpp2025_38211}.

\subsubsection{Time-Domain Signal Generation}
The discrete time-domain signal is generated uniformly across all \acp{TX} by applying the \ac{IFFT} and appending the \ac{CP}. For a generic frequency-domain payload $\mathbf{x}_F \in \{\mathbf{x}_{F,c}, \mathbf{x}_{F,s,1}, \dots, \mathbf{x}_{F,s,U}\}$, the corresponding time-domain signal is expressed as
\begin{align}
    \mathbf{x}_T &= \mathbf{A}_{cp} \mathbf{F}^H \mathbf{x}_F, \label{eq:time_domain_gen}
\end{align}
where $\mathbf{F}^H \in \mathbb{C}^{N \times N}$ denotes the \ac{IFFT} matrix, and $\mathbf{A}_{cp} \in \mathbb{R}^{(N+N_{cp}) \times N}$ is the \ac{CP} addition matrix that copies the last $N_{cp}$ samples to the beginning of the symbol to mitigate \ac{ISI}.

\subsection{Multipath Channel Model}
The \ac{UL} propagation environment exhibits frequency-selective multipath fading resulting from a rich scattering \ac{NLoS} environment. Let the subscript $\nu$ denote a generic \ac{TX} link, where $\nu = c$ corresponds to the \ac{CTX} and $\nu = s,u$ corresponds to the $u$-th \ac{STX}. For any individual link to the ISAC \ac{RX}, the baseband-equivalent time-varying \ac{CIR} is modeled as a tapped-delay line as follows: 
\begin{align}
    h_{T,\nu}(\tau) &= \sum_{l=0}^{L-1} \beta_{\nu,l} \delta(\tau - \tau_{\nu,l}), \label{eq:cir_model}
\end{align}
where $L$ denotes the number of resolvable multipath taps bounded by $N_{cp}$, $\beta_{\nu,l} \sim \mathcal{CN}(0, \sigma_h^2)$ represents the complex channel gain of the $l$-th propagation path, and $\tau_{\nu,l}$ specifies the continuous delay. The corresponding discrete time-domain channel vector is formally defined as $\mathbf{h}_{T,\nu} \in \mathbb{C}^{L \times 1}$. The frequency-domain channel response vector is explicitly denoted as $\mathbf{h}_{F,\nu} \in \mathbb{C}^{N \times 1}$, with its diagonalized representation defined as $\mathbf{H}_{F,\nu} = \text{diag}(\mathbf{h}_{F,\nu}) \in \mathbb{C}^{N \times N}$.

\subsection{Received Signal Model}
At the ISAC \ac{RX}, the aggregated received signal consists of the non-orthogonal superposition of the \ac{CTX} payload and the $U$ sensing sequences. Following analog-to-digital conversion, let $\mathbf{y}_T \in \mathbb{C}^{(N+N_{cp}) \times 1}$ denote the discrete time-domain received signal. The \ac{RX} isolates the core symbol by applying the \ac{CP} removal matrix $\mathbf{R}_{cp} \in \mathbb{R}^{N \times (N+N_{cp})}$, followed by the $N$-point FFT matrix $\mathbf{F}$. The resulting frequency-domain received signal vector, defined as $\mathbf{y}_F = \mathbf{F} \mathbf{R}_{cp} \mathbf{y}_T$, is mathematically modeled as
\begin{align}
    \mathbf{y}_F &= \mathbf{H}_{F,c} \mathbf{x}_{F,c} + \sum_{u=1}^{U} \mathbf{H}_{F,s,u} \mathbf{x}_{F,s,u} + \mathbf{w}_F, \label{eq:rx_superposition}
\end{align}
where $\mathbf{H}_{F,c} \in \mathbb{C}^{N \times N}$ is the frequency-domain channel matrix for the \ac{CTX}, $\mathbf{H}_{F,s,u} \in \mathbb{C}^{N \times N}$ denotes the frequency-domain channel matrix for the $u$-th sensing \ac{STX}, and $\mathbf{w}_F \sim \mathcal{CN}(\mathbf{0}, \sigma^2 \mathbf{I}_N)$ represents the frequency-domain \ac{AWGN} vector.

\section{Proposed Iterative Interference Cancellation Receivers}
\label{sec:proposed_method}
In the \ac{UL} \ac{ISAC}-\ac{NOMA} scenario, communication payloads and sensing pilots act as mutual interference. To resolve this, two specialized \ac{IIC} architectures are mathematically designed to decouple the signals.

\subsection{Joint IIC Architecture}
The joint \ac{IIC} strategy iteratively refines both the sensing and communication channels concurrently. Originally established in foundational single-\ac{STX} \ac{ISAC}-\ac{NOMA}~\cite{memisoglu2023csi}, this approach is mathematically adapted as a robust multi-\ac{STX} extension of standard joint detection.
Consequently, the proposed method is not restricted to non-nulling pilot schemes nor specific to the PS framework; its performance is evaluated across all considered pilot configurations. Let $q \in \{1, \dots, Q\}$ denote the iteration index. The iterative process commences by explicitly establishing the communication assumptions by initializing the \ac{DD} communication symbol vector as $\hat{\mathbf{x}}_{F,c,DD}^{(0)} = \mathbf{0}$ alongside the communication channel frequency response estimate as $\hat{\mathbf{h}}_{F,c}^{(0)} = \mathbf{0}$.

\subsubsection{Sensing Signal Recovery and Time-Domain CIR Separation}
At the $q$-th iteration, the \ac{RX} first mitigates the communication interference using the estimates from the previous iteration. The isolated sensing signal vector $\tilde{\mathbf{y}}_{F,s}^{(q)} \in \mathbb{C}^{N \times 1}$ is obtained by subtracting the reconstructed communication signal as
\begin{align}
    \tilde{\mathbf{y}}_{F,s}^{(q)} &= \mathbf{y}_F - \hat{\mathbf{H}}_{F,c}^{(q-1)} \hat{\mathbf{x}}_{F,c,DD}^{(q-1)}, \label{eq:sens_signal_update}
\end{align}
where $\hat{\mathbf{H}}_{F,c}^{(q-1)} = \text{diag}(\hat{\mathbf{h}}_{F,c}^{(q-1)})$ is the diagonal matrix of the estimated \ac{CTX} channel. Leveraging the phase-shifted frame structure, the composite sensing channel response is estimated via \ac{LS} estimation as 
\begin{align}
    \breve{\mathbf{h}}_{F,s,comp}^{(q)} &= \tilde{\mathbf{y}}_{F,s}^{(q)} \oslash \mathbf{x}_{F,r}. \label{eq:composite_freq_est}
\end{align}
The composite time-domain \ac{CIR} is then obtained via an $N$-point \ac{IFFT} matrix as
\begin{align}
    \breve{\mathbf{h}}_{T,s,comp}^{(q)} &= \mathbf{F}^H \breve{\mathbf{h}}_{F,s,comp}^{(q)}. \label{eq:composite_time_est}
\end{align}
Due to the \ac{TX}-specific phase shifts applied at each \ac{STX}, the \ac{CIR} of each sensing \ac{STX} is shifted into a distinct delay-domain window. The specific time-domain channel for the $u$-th \ac{STX}, denoted as $\tilde{\mathbf{h}}_{T,s,u}^{(q)}$, is extracted by applying a diagonal windowing matrix $\mathbf{W}_u \in \mathbb{R}^{N \times N}$, where the diagonal elements are $1$ for the time sample indices $n \in [(u-1)N_{cp}, uN_{cp}-1]$ and $0$ otherwise. The resulting channel is modeled as
\begin{align}
    \tilde{\mathbf{h}}_{T,s,u}^{(q)} &= \mathbf{W}_u \breve{\mathbf{h}}_{T,s,comp}^{(q)}. \label{eq:time_domain_windowing}
\end{align}
This time-domain gating filters out residual communication interference and out-of-band noise~\cite{ozdemir2007channel}. The refined frequency-domain sensing channel estimate for the $u$-th \ac{STX} is subsequently obtained by transforming $\tilde{\mathbf{h}}_{T,s,u}^{(q)}$ back to the frequency domain via an $N$-point FFT as
\begin{align}
    \hat{\mathbf{h}}_{F,s,u}^{(q)} &= \mathbf{F} \tilde{\mathbf{h}}_{T,s,u}^{(q)}. \label{eq:refined_sens_freq}
\end{align}

\subsubsection{Sensing Interference Cancellation}
Using the updated sensing channel estimates, the aggregate sensing interference is reconstructed and subtracted from the original received signal to isolate the communication band as
\begin{align}
    \tilde{\mathbf{y}}_{F,c}^{(q)} &= \mathbf{y}_F - \sum_{u=1}^{U} \hat{\mathbf{H}}_{F,s,u}^{(q)} \mathbf{x}_{F,s,u}, \label{eq:comm_signal_update}
\end{align}
where $\hat{\mathbf{H}}_{F,s,u}^{(q)} = \text{diag}(\hat{\mathbf{h}}_{F,s,u}^{(q)})$ is explicitly defined as the diagonalized representation of the refined sensing channel vector for the $u$-th scheduled \ac{STX}.

\subsubsection{Communication Channel Estimation}
An \ac{LS} estimate of the communication channel is first performed at the pilot subcarriers $\mathcal{K}_p$. Let $\mathbf{p}_{F,c} \in \mathbb{C}^{|\mathcal{K}_p| \times 1}$ denote the communication pilot symbols and $\mathbf{\Gamma}_p \in \mathbb{R}^{|\mathcal{K}_p| \times N}$ denote a sampling matrix that extracts the pilot indices. The initial pilot-based \ac{LS} channel estimate is computed as
\begin{align}
    \hat{\mathbf{h}}_{F,p}^{(q)} &= (\mathbf{\Gamma}_p \tilde{\mathbf{y}}_{F,c}^{(q)}) \oslash \mathbf{p}_{F,c}. \label{eq:comm_pilot_est}
\end{align}
To acquire the channel response over the entire bandwidth, including the data subcarriers $\mathcal{K}_d$, a \ac{DFT}-based interpolation is applied. By transforming the pilot-domain estimate into the delay domain, zero-padding the sequence to length $N$, and transforming it back to the frequency domain, the full-band communication channel vector is formally obtained as
\begin{align}
    \hat{\mathbf{h}}_{F,c}^{(q)} &= \mathbf{F} \left( \mathbf{T}_{zp} \mathbf{F}_p^H \hat{\mathbf{h}}_{F,p}^{(q)} \right), \label{eq:comm_fullband_est}
\end{align}
where $\mathbf{F}_p^H \in \mathbb{C}^{|\mathcal{K}_p| \times |\mathcal{K}_p|}$ denotes the unitary \ac{IFFT} matrix corresponding to the pilot dimension, and $\mathbf{T}_{zp} \in \mathbb{R}^{N \times |\mathcal{K}_p|}$ represents the zero-padding matrix that appends appropriate zeros to match the full FFT size.

\subsubsection{Data Detection and Decision-Directed Refinement}
Using the estimated communication channel, \ac{ZF} equalization is applied across the bandwidth to extract the soft data symbols as
\begin{align}
    \hat{\mathbf{x}}_{F,c}^{(q)} &= \tilde{\mathbf{y}}_{F,c}^{(q)} \oslash \hat{\mathbf{h}}_{F,c}^{(q)}. \label{eq:data_detect}
\end{align}
To reconstruct the \ac{DD} symbol vector over the entire bandwidth, the \ac{RX} projects the mapped data symbols and the known pilot symbols back into their respective subcarrier indices using the transpose of the sampling matrices $\mathbf{\Gamma}_d^T \in \mathbb{R}^{N \times |\mathcal{K}_d|}$ and $\mathbf{\Gamma}_p^T \in \mathbb{R}^{N \times |\mathcal{K}_p|}$ as 
\begin{align}
    \hat{\mathbf{x}}_{F,c,DD}^{(q)} &= \mathbf{\Gamma}_p^T \mathbf{p}_{F,c} + \mathbf{\Gamma}_d^T \mathcal{Q}\left( \mathbf{\Gamma}_d \hat{\mathbf{x}}_{F,c}^{(q)} \right). \label{eq:dd_reconstruction}
\end{align}
Using this \ac{DD} vector, the communication channel undergoes a final refinement step over the entire bandwidth given as
\begin{align}
    \hat{\mathbf{h}}_{F,c, DD}^{(q)} &= \tilde{\mathbf{y}}_{F,c}^{(q)} \oslash \hat{\mathbf{x}}_{F,c,DD}^{(q)}. \label{eq:dd_channel_refinement}
\end{align}
This refined channel estimate $\hat{\mathbf{h}}_{F,c, DD}^{(q)}$ and the updated \ac{DD} vector $\hat{\mathbf{x}}_{F,c,DD}^{(q)}$ are then fed back into \eqref{eq:sens_signal_update} for iteration $q+1$, completing the joint \ac{IIC} loop. The complete procedure for the joint \ac{IIC} architecture is formally summarized in Algorithm \ref{alg:joint_iic}.

\begin{algorithm}[t]
\caption{Proposed Joint IIC Architecture}
\label{alg:joint_iic}
\DontPrintSemicolon
\KwIn{$\mathbf{y}_F$, $\mathbf{x}_{F,r}$, $\mathbf{p}_{F,c}$, $Q$.}
\KwOut{$\hat{\mathbf{x}}_{F,c}^{(Q)}$, $\hat{\mathbf{h}}_{F,c}^{(Q)}$, $\{\hat{\mathbf{h}}_{F,s,u}^{(Q)}\}_{u=1}^U$.}
\textbf{Initialization: } Set $\hat{\mathbf{h}}_{F,c}^{(0)} \leftarrow \mathbf{0}$, $\hat{\mathbf{x}}_{F,c,DD}^{(0)} \leftarrow \mathbf{0}$\;
\For{$q \leftarrow 1$ \KwTo $Q$}{
    \vspace{2pt}
    \textbf{1. Sensing Signal Recovery}\;
    $\tilde{\mathbf{y}}_{F,s}^{(q)} \leftarrow \mathbf{y}_F - \hat{\mathbf{H}}_{F,c}^{(q-1)} \hat{\mathbf{x}}_{F,c,DD}^{(q-1)}$ \hfill \eqref{eq:sens_signal_update}\;
    $\breve{\mathbf{h}}_{F,s,comp}^{(q)} \leftarrow \tilde{\mathbf{y}}_{F,s}^{(q)} \oslash \mathbf{x}_{F,r}$ \hfill \eqref{eq:composite_freq_est}\;
    $\breve{\mathbf{h}}_{T,s,comp}^{(q)} \leftarrow \mathbf{F}^H \breve{\mathbf{h}}_{F,s,comp}^{(q)}$ \hfill \eqref{eq:composite_time_est}\;
    \For{$u \leftarrow 1$ \KwTo $U$}{
        $\tilde{\mathbf{h}}_{T,s,u}^{(q)} \leftarrow \mathbf{W}_u \breve{\mathbf{h}}_{T,s,comp}^{(q)}$ \hfill \eqref{eq:time_domain_windowing}\;
        $\hat{\mathbf{h}}_{F,s,u}^{(q)} \leftarrow \mathbf{F} \tilde{\mathbf{h}}_{T,s,u}^{(q)}$ \hfill \eqref{eq:refined_sens_freq}\;
        }
    \vspace{2pt}
    \textbf{2. Sensing Interference Cancellation}\;
    $\tilde{\mathbf{y}}_{F,c}^{(q)} \leftarrow \mathbf{y}_F - \sum_{u=1}^{U} \hat{\mathbf{H}}_{F,s,u}^{(q)} \mathbf{x}_{F,s,u}$ \hfill \eqref{eq:comm_signal_update}\;
   \vspace{2pt}
    \textbf{3. Communication Channel Estimation}\;
    $\hat{\mathbf{h}}_{F,p}^{(q)} \leftarrow (\mathbf{\Gamma}_p \tilde{\mathbf{y}}_{F,c}^{(q)}) \oslash \mathbf{p}_{F,c}$ \hfill \eqref{eq:comm_pilot_est}\;
    $\hat{\mathbf{h}}_{F,c}^{(q)} \leftarrow \mathbf{F} \left( \mathbf{T}_{zp} \mathbf{F}_p^H \hat{\mathbf{h}}_{F,p}^{(q)} \right)$ \hfill \eqref{eq:comm_fullband_est}\;
    \vspace{2pt}
    \textbf{4. Data Detection and DD Refinement}\;
    $\hat{\mathbf{x}}_{F,c}^{(q)} \leftarrow \tilde{\mathbf{y}}_{F,c}^{(q)} \oslash \hat{\mathbf{h}}_{F,c}^{(q)}$ \hfill \eqref{eq:data_detect}\;
    $\hat{\mathbf{x}}_{F,c,DD}^{(q)} \leftarrow \mathbf{\Gamma}_p^T \mathbf{p}_{F,c} + \mathbf{\Gamma}_d^T \mathcal{Q}\left( \mathbf{\Gamma}_d \hat{\mathbf{x}}_{F,c}^{(q)} \right)$ \hfill \eqref{eq:dd_reconstruction}\;
    $\hat{\mathbf{h}}_{F,c, DD}^{(q)} \leftarrow \tilde{\mathbf{y}}_{F,c}^{(q)} \oslash \hat{\mathbf{x}}_{F,c,DD}^{(q)}$ \hfill \eqref{eq:dd_channel_refinement}\;
    Update $\hat{\mathbf{h}}_{F,c}^{(q)} \leftarrow \hat{\mathbf{h}}_{F,c,DD}^{(q)}$ for the next iteration\;
}
\Return $\hat{\mathbf{x}}_{F,c}^{(Q)}, \hat{\mathbf{h}}_{F,c}^{(Q)}, \{\hat{\mathbf{h}}_{F,s,u}^{(Q)}\}_{u=1}^U$\;
\end{algorithm}

\subsection{Sequential IIC Architecture}
The sequential \ac{IIC} architecture operates on a mathematically distinct flow, systematically locking the initial communication channel estimate during early iterations to accelerate convergence and reduce computational complexity~\cite{zheng2025csi}. The algorithm requires an initial communication channel frequency response estimate, formally denoted as $\hat{\mathbf{h}}_{F,c}^{(0)}$, which is obtained via standard \ac{LS} estimation at the pilot indices followed by \ac{DFT}-based interpolation, as explicitly formulated in~\eqref{eq:comm_pilot_est} and~\eqref{eq:comm_fullband_est}.

\subsubsection{Sequential Cancellation for Iteration Indices $q \le Q_1$}
Establishing the equality $\hat{\mathbf{H}}_{F,c}^{(q)} = \hat{\mathbf{H}}_{F,c}^{(0)}$, the \ac{RX} exclusively updates the sensing channels and the detected data symbols during the first $Q_1$ iterations. In the $q$-th iteration, the isolated sensing signal vector, defined as $\tilde{\mathbf{y}}_{F,s}^{(q)} \in \mathbb{C}^{N \times 1}$, is obtained by subtracting the communication interference utilizing the static channel estimate as follows: 
\begin{align}
    \tilde{\mathbf{y}}_{F,s}^{(q)} &= \mathbf{y}_F - \hat{\mathbf{H}}_{F,c}^{(0)} \hat{\mathbf{x}}_{F,c,DD}^{(q-1)}. \label{eq:seq_sens_signal}
\end{align}
Utilizing the procedural consistency established in the joint \ac{IIC} approach, the time-domain composite sensing channel is extracted via an \ac{IFFT}, evaluated through the operations in \eqref{eq:composite_freq_est} and \eqref{eq:composite_time_est}. The specific \ac{CIR} for the $u$-th scheduled \ac{STX}, denoted as $\tilde{\mathbf{h}}_{T,s,u}^{(q)}$, is isolated utilizing the delay-domain windowing matrix $\mathbf{W}_u$, as defined in \eqref{eq:time_domain_windowing}. An $N$-point FFT subsequently yields the refined frequency-domain sensing channel vector $\hat{\mathbf{h}}_{F,s,u}^{(q)}$, mirroring the result in \eqref{eq:refined_sens_freq}.

Subsequently, the updated sensing channels are utilized to reconstruct and cancel the aggregate sensing interference from the received signal, yielding the isolated communication signal $\tilde{\mathbf{y}}_{F,c}^{(q)}$ as previously formulated in \eqref{eq:comm_signal_update}. The communication data symbols are then directly detected via \ac{ZF} equalization utilizing the fixed initial communication channel and given as
\begin{align}
    \hat{\mathbf{x}}_{F,c}^{(q)} &= \tilde{\mathbf{y}}_{F,c}^{(q)} \oslash \hat{\mathbf{h}}_{F,c}^{(0)}. \label{eq:seq_data_detect}
\end{align}
Subsequently, $\hat{\mathbf{x}}_{F,c,DD}^{(q)}$ is reconstructed by mapping the data subcarriers of $\hat{\mathbf{x}}_{F,c}^{(q)}$ to the corresponding constellation points, following the mapping operator previously defined in \eqref{eq:dd_reconstruction}.

\subsubsection{Decision-Directed Refinement for Iteration Indices $Q_1 < q \le Q_1+Q_2$}
Once the sensing interference is structurally mitigated, the reconstructed \ac{DD} symbol vector at the conclusion of the first phase, formally denoted as $\hat{\mathbf{x}}_{F,c,DD}^{(Q_1)}$, transitions to a refined state. In the subsequent $Q_2$ iterations, the algorithm unfreezes the communication channel and performs a joint \ac{DD} refinement.

For iteration indices exceeding $Q_1$, the sensing signal is updated utilizing the dynamic communication channel estimate, in accordance with \eqref{eq:sens_signal_update}. After extracting the refined sensing channels via the sequence from \eqref{eq:composite_freq_est} through \eqref{eq:refined_sens_freq} and subsequently updating the isolated communication signal $\tilde{\mathbf{y}}_{F,c}^{(q)}$ via \eqref{eq:comm_signal_update}, the communication channel is re-estimated utilizing the \ac{DD} symbols. To suppress residual noise, a \ac{DFT}-based noise cancellation operation is applied. The time-domain communication channel is first estimated as
\begin{align}
    \tilde{\mathbf{h}}_{T,c}^{(q)} &= \mathbf{F}^H \left( \tilde{\mathbf{y}}_{F,c}^{(q)} \oslash \hat{\mathbf{x}}_{F,c,DD}^{(q-1)} \right). \label{eq:dd_time_channel}
\end{align}
The updated frequency-domain communication channel is then obtained by applying a truncation window matrix $\mathbf{W}_c \in \mathbb{R}^{N \times N}$, which isolates the samples corresponding to the \ac{CP} duration, followed by the application of an FFT
\begin{align}
    \hat{\mathbf{h}}_{F,c}^{(q)} &= \mathbf{F} \left( \mathbf{W}_c \tilde{\mathbf{h}}_{T,c}^{(q)} \right). \label{eq:dd_freq_channel}
\end{align}
Finally, the soft data symbols are detected using the refined channel $\hat{\mathbf{h}}_{F,c}^{(q)}$, as formulated in \eqref{eq:data_detect}, and the \ac{DD} vector $\hat{\mathbf{x}}_{F,c,DD}^{(q)}$ is reconstructed according to \eqref{eq:dd_reconstruction} to serve as the updated reference for the next iteration. The complete procedure for the sequential \ac{IIC} architecture is formally summarized in Algorithm \ref{alg:seq_iic}.

\begin{algorithm}[t]
\caption{Proposed Sequential IIC Architecture}
\label{alg:seq_iic}
\DontPrintSemicolon
\KwIn{$\mathbf{y}_F$, $\mathbf{x}_{F,r}$, $\mathbf{x}_{F,s,u}$, $\mathbf{p}_{F,c}$, $Q_1$, $Q_2$.}
\KwOut{$\hat{\mathbf{x}}_{F,c}^{(Q_1+Q_2)}$, $\hat{\mathbf{h}}_{F,c}^{(Q_1+Q_2)}$, $\{\hat{\mathbf{h}}_{F,s,u}^{(Q_1+Q_2)}\}_{u=1}^U$.}
\textbf{Initialization:}\;
$\hat{\mathbf{h}}_{F,p}^{(0)} \leftarrow (\mathbf{\Gamma}_p \mathbf{y}_F) \oslash \mathbf{p}_{F,c}$ \hfill \eqref{eq:comm_pilot_est}\;
$\hat{\mathbf{h}}_{F,c}^{(0)} \leftarrow \mathbf{F} \left( \mathbf{T}_{zp} \mathbf{F}_p^H \hat{\mathbf{h}}_{F,p}^{(0)} \right)$ \hfill \eqref{eq:comm_fullband_est}\;
$\hat{\mathbf{x}}_{F,c,DD}^{(0)} \leftarrow \mathbf{0}$\;
\For{$q \leftarrow 1$ \KwTo $Q_1 + Q_2$}{
    \vspace{2pt}
    \textbf{1. Sensing Signal Recovery}\;
    \eIf{$q \le Q_1$}{
        $\tilde{\mathbf{y}}_{F,s}^{(q)} \leftarrow \mathbf{y}_F - \hat{\mathbf{H}}_{F,c}^{(0)} \hat{\mathbf{x}}_{F,c,DD}^{(q-1)}$ \hfill \eqref{eq:seq_sens_signal}\;
    }{
        $\tilde{\mathbf{y}}_{F,s}^{(q)} \leftarrow \mathbf{y}_F - \hat{\mathbf{H}}_{F,c}^{(q-1)} \hat{\mathbf{x}}_{F,c,DD}^{(q-1)}$ \hfill \eqref{eq:sens_signal_update}\;
    }
    $\breve{\mathbf{h}}_{F,s,comp}^{(q)} \leftarrow \tilde{\mathbf{y}}_{F,s}^{(q)} \oslash \mathbf{x}_{F,r}$ \hfill \eqref{eq:composite_freq_est}\;
    $\breve{\mathbf{h}}_{T,s,comp}^{(q)} \leftarrow \mathbf{F}^H \breve{\mathbf{h}}_{F,s,comp}^{(q)}$ \hfill \eqref{eq:composite_time_est}\;
    \For{$u \leftarrow 1$ \KwTo $U$}{
        $\tilde{\mathbf{h}}_{T,s,u}^{(q)} \leftarrow \mathbf{W}_u \breve{\mathbf{h}}_{T,s,comp}^{(q)}$ \hfill \eqref{eq:time_domain_windowing}\;
        $\hat{\mathbf{h}}_{F,s,u}^{(q)} \leftarrow \mathbf{F} \tilde{\mathbf{h}}_{T,s,u}^{(q)}$ \hfill \eqref{eq:refined_sens_freq}\;
    }
    \vspace{2pt}
    \textbf{2. Sensing Interference Cancellation}\;
    $\tilde{\mathbf{y}}_{F,c}^{(q)} \leftarrow \mathbf{y}_F - \sum_{u=1}^{U} \hat{\mathbf{H}}_{F,s,u}^{(q)} \mathbf{x}_{F,s,u}$ \hfill \eqref{eq:comm_signal_update}\;
    \vspace{2pt}
    \textbf{3. Communication Channel Refinement}\;
    \If{$q > Q_1$}{
        $\tilde{\mathbf{h}}_{T,c}^{(q)} \leftarrow \mathbf{F}^H \left( \tilde{\mathbf{y}}_{F,c}^{(q)} \oslash \hat{\mathbf{x}}_{F,c,DD}^{(q-1)} \right)$ \hfill \eqref{eq:dd_time_channel}\;
        $\hat{\mathbf{h}}_{F,c}^{(q)} \leftarrow \mathbf{F} \left( \mathbf{W}_c \tilde{\mathbf{h}}_{T,c}^{(q)} \right)$ \hfill \eqref{eq:dd_freq_channel}\;
    }
    \vspace{2pt}
    \textbf{4. Data Detection and DD Refinement}\;
    \eIf{$q \le Q_1$}{
        $\hat{\mathbf{x}}_{F,c}^{(q)} \leftarrow \tilde{\mathbf{y}}_{F,c}^{(q)} \oslash \hat{\mathbf{h}}_{F,c}^{(0)}$ \hfill \eqref{eq:seq_data_detect}\;
    }{
        $\hat{\mathbf{x}}_{F,c}^{(q)} \leftarrow \tilde{\mathbf{y}}_{F,c}^{(q)} \oslash \hat{\mathbf{h}}_{F,c}^{(q)}$ \hfill \eqref{eq:data_detect}\;
    }
    $\hat{\mathbf{x}}_{F,c,DD}^{(q)} \leftarrow \mathbf{\Gamma}_p^T \mathbf{p}_{F,c} + \mathbf{\Gamma}_d^T \mathcal{Q}\left( \mathbf{\Gamma}_d \hat{\mathbf{x}}_{F,c}^{(q)} \right)$ \hfill \eqref{eq:dd_reconstruction}\;
}
\Return $\hat{\mathbf{x}}_{F,c}^{(Q_1+Q_2)}, \hat{\mathbf{h}}_{F,c}^{(Q_1+Q_2)}, \{\hat{\mathbf{h}}_{F,s,u}^{(Q_1+Q_2)}\}_{u=1}^U$\;
\end{algorithm}

\begin{table*}[t]
\centering
\caption{Computational Complexity of ISAC-NOMA Architectures}
\label{tab:complexity}
\renewcommand{\arraystretch}{1.5}
\begin{tabular}{|l|l|c|c|c|}
\hline
\begin{tabular}[c]{@{}l@{}}\textbf{Architecture} \\ (IIC Strategy)\end{tabular} & \textbf{Processing Node} & \textbf{Real Additions} & \textbf{Real Multiplications} & \textbf{Asymptotic Scaling} \\ \hline\hline
\multirow{2}{*}{\begin{tabular}[c]{@{}l@{}}\textbf{PS-ISAC-NOMA} \\ (Joint, $Q$ iterations)\end{tabular}} 
& STX Side  
& $U(C_A + 2N)$ 
& $U(C_M + 4N)$ 
& $\mathcal{O}(U N \log_2 N)$ \\ \cline{2-5} 
& ISAC RX Side  
& $Q\big[(U+3) C_A + 4(U+3)N\big]$ 
& $Q\big[(U+3) C_M + 4(U+5)N\big]$ 
& $\mathcal{O}\big(Q(U+3) N \log_2 N\big)$ \\ \hline
\multirow{2}{*}{\begin{tabular}[c]{@{}l@{}}\textbf{CI-ISAC-NOMA} \\ (Joint, $Q$ iterations)\end{tabular}} 
& STX Side  
& $U \cdot C_A$ 
& $U \cdot C_M$ 
& $\mathcal{O}(U N \log_2 N)$ \\ \cline{2-5} 
& ISAC RX Side  
& $Q\big[(2U+2) C_A + 4(U+3)N\big]$ 
& $Q\big[(2U+2) C_M + 4(U+5)N\big]$ 
& $\mathcal{O}\big(Q(2U+2) N \log_2 N\big)$ \\ \hline
\multirow{2}{*}{\begin{tabular}[c]{@{}l@{}}\textbf{PSN-ISAC-NOMA} \\ (Sequential, \\
$Q_1 + Q_2$ iterations)\end{tabular}} 
& STX Side  
& $U(C_A + 2N)$ 
& $U(C_M + 4N)$ 
& $\mathcal{O}(U N \log_2 N)$ \\ \cline{2-5} 
& ISAC RX Side  
& \begin{tabular}[c]{@{}c@{}}
$Q_1\big[(U+1)C_A + 4(U+1)N\big] + $\\ $Q_2\big[(U+3)C_A + 4(U+3)N\big]$
\end{tabular} 
& \begin{tabular}[c]{@{}c@{}}
$Q_1\big[(U+1)C_M + 4(U+2)N\big] + $
\\ $Q_2\big[(U+3)C_M + 4(U+5)N\big]$
\end{tabular} 
& \begin{tabular}[c]{@{}c@{}}
 $\mathcal{O}\big([Q_1(U+1) + $\\ $ Q_2(U+3)] N \log_2 N\big)$
\end{tabular}  \\ \hline
\multirow{2}{*}{\begin{tabular}[c]{@{}l@{}}\textbf{CIN-ISAC-NOMA} \\ (Sequential, \\ $Q_1 + Q_2$ iterations)
\end{tabular}} 
& STX Side  
& $U \cdot C_A$ 
& $U \cdot C_M$ 
& $\mathcal{O}(U N \log_2 N)$ \\ \cline{2-5} 
& ISAC RX Side  
& \begin{tabular}[c]{@{}c@{}}
$Q_1\big[(2U)C_A + 4(U+1)N\big] + $ \\ $Q_2\big[(2U+2)C_A + 4(U+3)N\big]$ 
\end{tabular} 
& \begin{tabular}[c]{@{}c@{}}
 $Q_1\big[(2U)C_M + 4(U+2)N\big] + $ \\  $Q_2\big[(2U+2)C_M + 4(U+5)N\big]$ 
\end{tabular} 
& \begin{tabular}[c]{@{}c@{}}
 $\mathcal{O}\big([Q_1(2U) + $ \\  $ Q_2(2U+2)] N \log_2 N\big)$ 
\end{tabular}  \\ \hline
\end{tabular}
\end{table*}

\section{Computational Complexity Analysis}
\label{sec:complexity}
In this section, the computational complexity of the proposed phase-shifted architectures is mathematically evaluated and compared against the interleaved baselines. Complexity is quantified by the total number of real additions and real multiplications. A standard complex addition requires $2$ real additions, while a standard complex multiplication requires $4$ real multiplications and $2$ real additions, as established in~\cite{6476061}. For element-wise vector division, multiplication by a precomputed inverse is assumed, equating the hardware cost to a standard complex multiplication~\cite{proakis2001digital}.
In baseband processing, the dominant burden arises from the FFT and \ac{IFFT} operations. The baseline complexity of an $N$-point unitary transform is formally defined as $C_A = 3N \log_2 N - 3N + 4$ real additions and $C_M = N \log_2 N - 3N + 4$ real multiplications~\cite{sorensen1986computing}. Table \ref{tab:complexity} provides a comprehensive summary of the arithmetic operations and asymptotic scaling required by the STX and ISAC \ac{RX}.

\subsection{STX-Side Complexity}
At the \ac{STX} side, generating the time-domain signal via \eqref{eq:time_domain_gen} requires an $N$-point \ac{IFFT} for all evaluated architectures. Consequently, for $U$ scheduled \acp{STX}, the baseline \ac{STX} complexity is $U \cdot C_A$ additions and $U \cdot C_M$ multiplications.

For the proposed pilot designs, an additional \ac{STX}-specific phase shift mapping is applied. This element-wise cyclic phase rotation requires multiplying the base sequence by the diagonal matrix $\mathbf{P}_u(\theta_u)$ defined in \eqref{eq:phase_shift_matrix}, utilizing exactly $4N$ real multiplications and $2N$ real additions per \ac{STX}. Thus, they incur a total of $U(C_A + 2N)$ additions and $U(C_M + 4N)$ multiplications. This introduces only a marginal linear overhead, keeping the \ac{STX}-side complexity tightly bounded at $\mathcal{O}(UN \log_2 N)$.

\subsection{Receiver-Side Complexity: Joint IIC Architecture}
At the ISAC \ac{RX}, the joint \ac{IIC} algorithm disentangles the \ac{NOMA} superposition. After the initial FFT common to all methods, the exact complexity per iteration $q$ is defined by tracing the mathematical operators given as
\begin{itemize}
    \item Sensing Recovery: Reconstructing and subtracting the communication signal in \eqref{eq:sens_signal_update} requires $4N$ real multiplications and $4N$ real additions. The composite channel extraction via Hadamard division in \eqref{eq:composite_freq_est} adds $4N$ real multiplications and $2N$ real additions.
    \item Composite Transformation: The extraction of the composite time-domain response in \eqref{eq:composite_time_est} requires $1$ \ac{IFFT}.  Transforming the separated \acp{CIR} back to the frequency domain via \eqref{eq:refined_sens_freq} demands $U$ individual FFTs.
    \item Communication Isolation: Canceling the aggregate sensing interference in \eqref{eq:comm_signal_update} necessitates $U$ complex vector multiplications and $U$ complex vector additions, summing to $4UN$ real multiplications and $4UN$ real additions.
    \item Channel Estimation \& Data Detection: The pilot-based estimation and \ac{DFT} interpolation in \eqref{eq:comm_pilot_est} and \eqref{eq:comm_fullband_est} conservatively require $1$ \ac{IFFT}, $1$ FFT, plus $4N$ real multiplications and $2N$ real additions. The data equalization via \eqref{eq:data_detect} incurs $4N$ real multiplications and $2N$ real additions. The \ac{DD} refinement in \eqref{eq:dd_channel_refinement} adds a final $4N$ real multiplications and $2N$ real additions.
\end{itemize}
Aggregating these operators, the joint \ac{IIC} requires $U+3$ total transforms and an exact linear overhead of $4(U+5)N$ multiplications and $4(U+3)N$ additions per iteration. 

\subsection{Receiver-Side Complexity: Sequential IIC Architecture}
The sequential \ac{IIC} architecture, explicitly adapted for the PSN and CIN spectral nulling variants, significantly reduces the initial hardware burden by systematically fixing the initial communication channel.
\begin{itemize}
    \item Phase 1, comprising the cancellation stage where $q \le Q_1$: The \ac{CTX} channel is fixed. Updating the sensing signal via \eqref{eq:seq_sens_signal} requires $4N$ multiplications and $4N$ additions. Time-domain extraction mirrors the joint method requiring $U+1$ transforms. Communication isolation via \eqref{eq:comm_signal_update} requires $4UN$ multiplications and $4UN$ additions. Crucially, by executing data detection directly via \eqref{eq:seq_data_detect}, the algorithm entirely bypasses the \ac{DFT}-based channel refinement penalty. This phase requires only $U+1$ transforms, alongside $4(U+2)N$ multiplications and $4(U+1)N$ additions per iteration.
    \item Phase 2, explicitly the \ac{DD} refinement stage where $q > Q_1$: The algorithm unfreezes the \ac{CTX} channel. Applying the \ac{DFT}-based noise cancellation in \eqref{eq:dd_time_channel} and \eqref{eq:dd_freq_channel} reintroduces $1$ \ac{IFFT}, $1$ FFT, plus $4N$ multiplications and $2N$ additions, aligning the transform count with the joint \ac{IIC} at $U+3$ transforms per iteration.
\end{itemize}
By explicitly accounting for both the $\mathcal{O}(Q U N \log_2 N)$ transform costs and the $\mathcal{O}(Q U N)$ element-wise operations derived from the \ac{IIC} equations, the hardware efficiency of the phase-shifted architectures is formally quantified.

\section{Simulation Results}
\label{sec:simulation_results}
In this section, the performance of the proposed PS-\ac{ISAC}-\ac{NOMA} and PSN-\ac{ISAC}-\ac{NOMA} frameworks is rigorously evaluated against the CI and CIN baselines. Simulations are conducted over frequency-selective Rayleigh fading channels with $L=8$ resolvable taps. The system employs an \ac{OFDM} architecture with $N=512$ subcarriers and $N_{cp}=8$. Unless otherwise specified, the default parameters are set to $U=4$ \acp{STX} and a \ac{CTX} modulation order of $M=4$.

To quantify the interference cancellation efficiency, the system is evaluated under a normalized average transmission power establishing the \ac{SNR} explicitly as $\text{SNR} = 1/\sigma^2$. The sensing \ac{NMSE} averaged across the entire multi-STX environment is mathematically formulated as 
\begin{equation}
    \text{Sensing NMSE} = \frac{1}{U} \sum_{u=1}^{U} \frac{\mathbb{E}[\|\mathbf{h}_{F,s,u} - \hat{\mathbf{h}}_{F,s,u}\|_2^2]}{\mathbb{E}[\|\mathbf{h}_{F,s,u}\|_2^2]}.
\end{equation}
Similarly, the communication \ac{NMSE} isolating the \ac{CTX} link is explicitly defined as 
\begin{equation}
    \text{Communication NMSE} = \frac{\mathbb{E}[\|\mathbf{h}_{F,c} - \hat{\mathbf{h}}_{F,c}\|_2^2]}{\mathbb{E}[\|\mathbf{h}_{F,c}\|_2^2]}.
\end{equation}

\subsection{IIC Strategy and Convergence Analysis}
\begin{figure}
\centering
\includegraphics[width=0.47\textwidth]{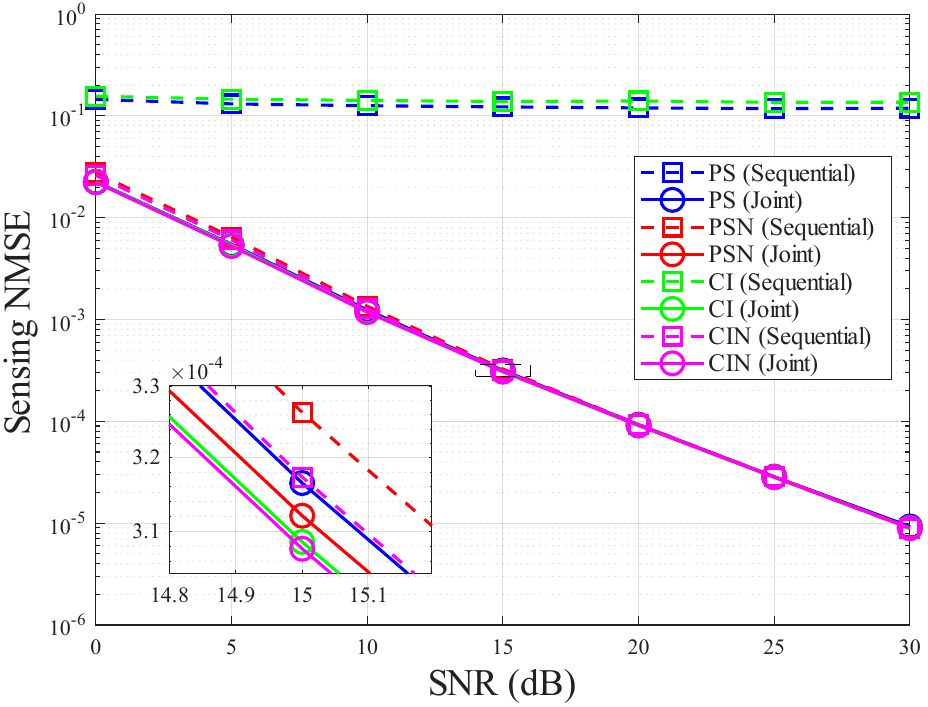}
\caption{Outlier analysis of the \ac{IIC} architectures evaluating the sensing \ac{NMSE} versus the \ac{SNR}, specifically for a network density of $U=4$ scheduled \acp{STX} operating under a fixed modulation order of $M=4$. }
\label{fig:mic_sensing_outlier}
\end{figure}

The sensitivity of the \ac{RX} architectures to spectral overlap is first evaluated to mathematically determine the optimal \ac{IIC} strategy. As shown in the outlier analysis in Fig.~\ref{fig:mic_sensing_outlier}, the joint IIC strategy empirically converges for all schemes by concurrently updating communication and sensing channel estimates. Conversely, the sequential IIC strategy is incompatible with non-nulling architectures. Without spectral puncturing, the non-nulling sensing signals overlap with \ac{CTX} pilot tones, causing a biased initial communication estimate $\hat{\mathbf{h}}_{F,c}^{(0)}$. Freezing this inaccurate estimate propagates errors through subsequent iterations, preventing the \ac{RX} from reaching the expected error floor. Consequently, for the PS and CI variants, the joint IIC architecture is exclusively employed.

\begin{figure}
\centering
\subfigure[]
{\includegraphics[width=0.47\textwidth]{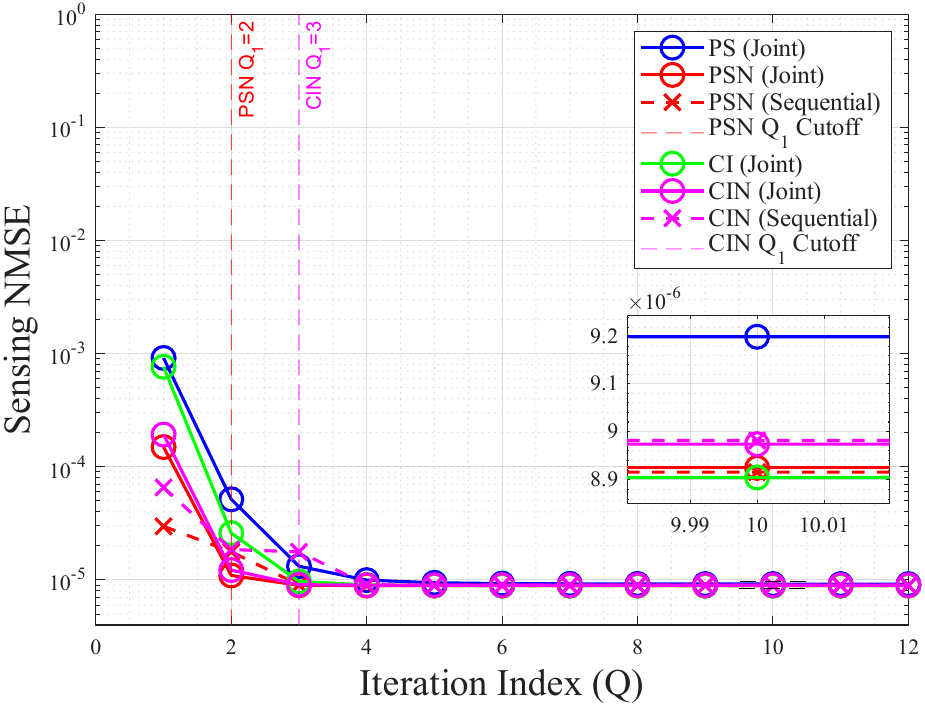}}
\subfigure[]
{\includegraphics[width=0.47\textwidth]{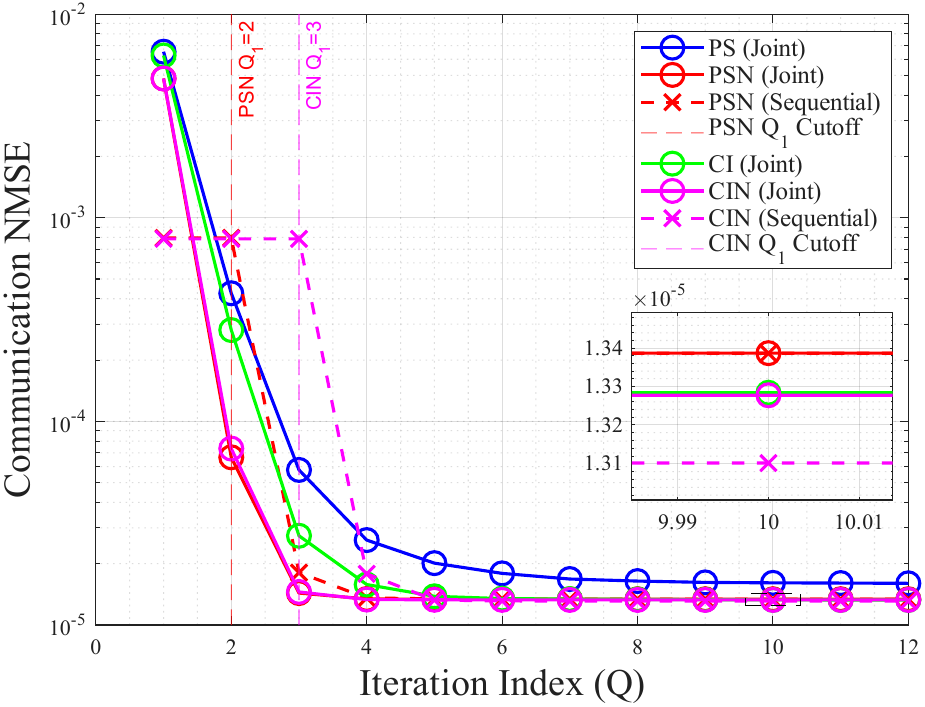}}
\subfigure[]
{\includegraphics[width=0.47\textwidth]{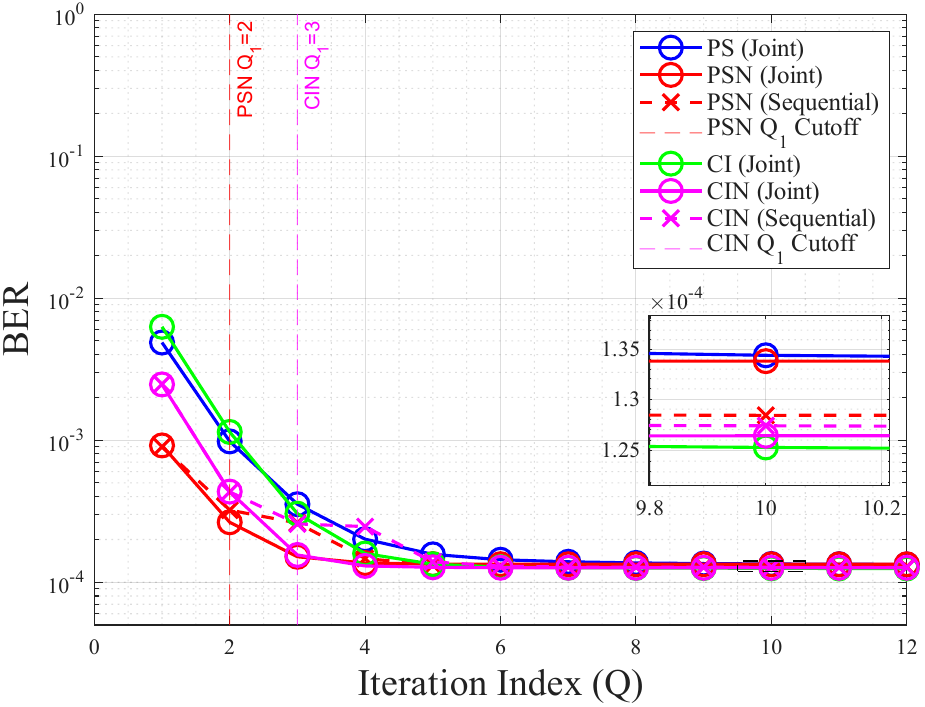}}
\caption{Convergence analysis of the \ac{IIC} architectures for $U=4$ and $M=4$: explicitly subfigure (a) illustrating sensing \ac{NMSE} convergence, subfigure (b) depicting communication \ac{NMSE} convergence, and subfigure (c) showing \ac{BER} convergence. The iteration parameters $Q_1 = 3$, $Q_2 = 4$, and $Q = 7$ are selected based on the common maximum iterations required to satisfy a relative performance improvement threshold of $10^{-4}$ across all evaluated schemes.}
\label{fig:moi_metrics}
\end{figure}

Following the formal establishment of viable \ac{RX} pairings, the subsequent convergence trajectories illustrated in Fig.~\ref{fig:moi_metrics} are rigorously analyzed. For the nulling-based frameworks, spectral puncturing ensures an interference-free initial communication estimate, allowing both joint and sequential architectures to converge to identical floors. While the sensing \ac{NMSE} and \ac{BER} follow similar convergence paths, as shown in Fig.~\ref{fig:moi_metrics}(a) and Fig.~\ref{fig:moi_metrics}(c), the communication \ac{NMSE} in Fig.~\ref{fig:moi_metrics}(b) exhibits a distinct trajectory under the sequential strategy. This is due to the deliberate fixing of the communication state during early loops, followed by \ac{DD} refinement once multi-\ac{STX} interference is sufficiently suppressed.

To formalize the iteration limits, convergence is explicitly defined as a relative performance improvement of less than $10^{-4}$ between consecutive steps. Based on this criterion, the sequential phase reaches saturation, explicitly denoted as the $Q_1$ cutoff, in two iterations for PSN and three for CIN. To ensure a fair comparative baseline, uniform parameters of $Q_1=3$ and $Q_2=4$ are established for all nulling schemes. Because the sequential approach provides faster early-stage convergence and significantly lower computational complexity, it is selected as the default strategy for PSN and CIN. For consistency, the joint IIC utilized by the non-nulling variants is also executed for $Q=7$ iterations. 
While the convergence trajectories in Fig.~\ref{fig:moi_metrics} are explicitly illustrated for the baseline configuration of $M=4$ and $U=4$, the optimal iteration parameters for all other simulated scenarios were determined using the same rigorous methodology.

\subsection{Performance and Scalability Analysis}
\label{subsec:performance_and_scaling}

To evaluate the robustness of the adapted \ac{IIC} \acp{RX}, comprehensive evaluations are conducted by systematically varying the \ac{CTX} modulation order $M$ and the number of scheduled \acp{STX} $U$. The performance of the evaluated frameworks is benchmarked against theoretical lower bounds (LBs): The PS LB and CI LB represent orthogonal sensing architectures operating without \ac{NOMA} interference, while the CO LB represents the interference-free communication-only scheme.

\begin{figure}
\centering
\subfigure[]
{\includegraphics[width=0.47\textwidth]{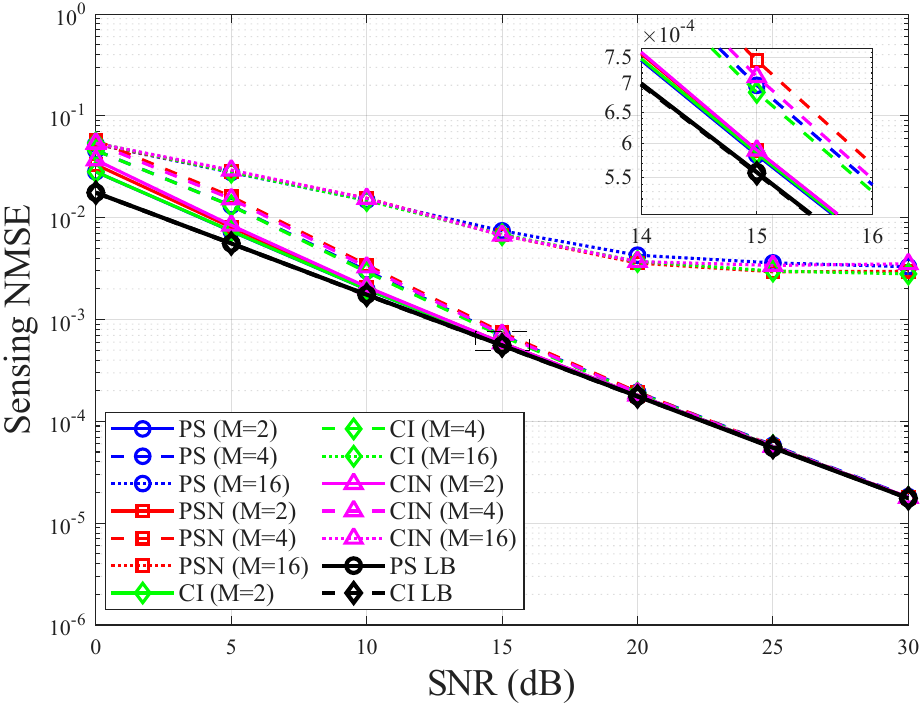}}
\subfigure[]
{\includegraphics[width=0.47\textwidth]{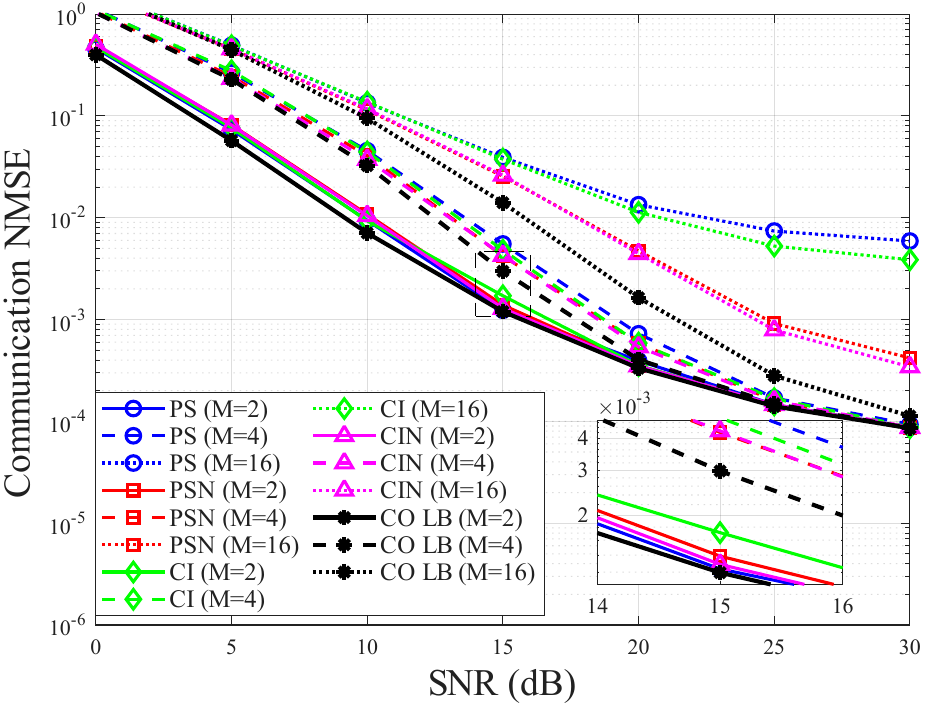}}
\subfigure[]
{\includegraphics[width=0.47\textwidth]{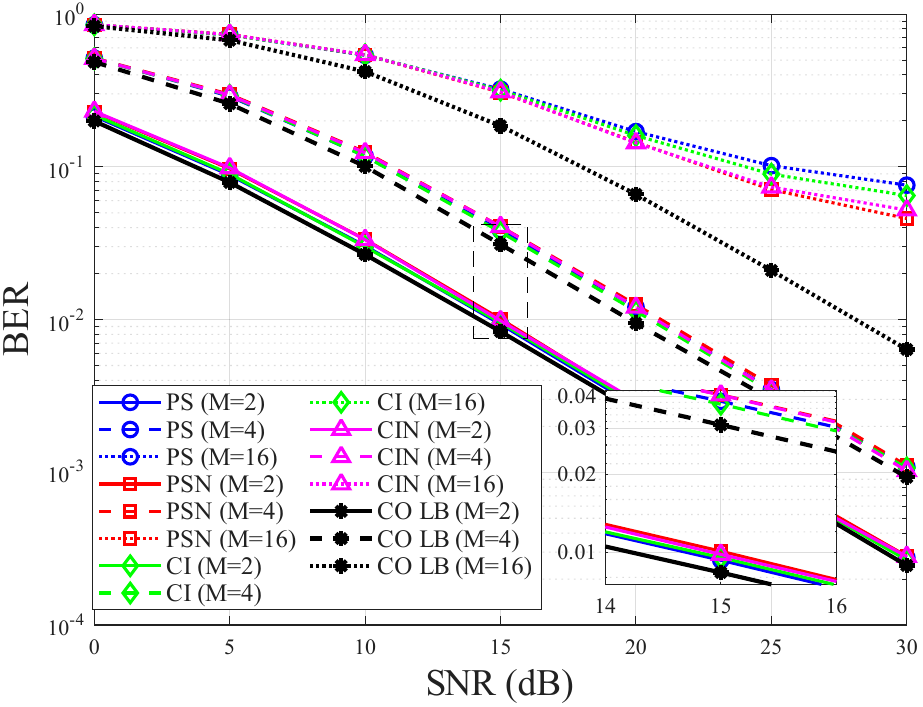}}
\caption{Robustness analysis of the ISAC-NOMA frameworks against the \ac{CTX} modulation order for $M \in \{2, 4, 16\}$ and $U=4$: subfigure (a) illustrating sensing \ac{NMSE}, subfigure (b) depicting communication \ac{NMSE}, and subfigure (c) showing the \ac{BER}. The theoretical limits, explicitly defined as the CO LB, PS LB, and CI LB, represent the communication-only scenario as well as the phase-shifted sensing-only, denoted as PS-ISAC, and interleaved sensing-only, denoted as CI-ISAC, baseline configurations.}
\label{fig:mkf_metrics}
\end{figure}

System robustness is first evaluated by sweeping the \ac{CTX} modulation order across $M \in \{2, 4, 16\}$ for a fixed density of $U=4$ sensing \acp{STX}. As illustrated in the sensing \ac{NMSE} trajectories in Fig.~\ref{fig:mkf_metrics}(a), all \ac{NOMA} architectures closely track their respective benchmarks for $M=2$ and $M=4$. Specifically, beyond an \ac{SNR} of $20$ dB, the proposed and conventional schemes perfectly overlap with the PS LB and CI LB. However, transitioning to the extreme dynamic range of $M=16$ prevents the \ac{IIC} \acp{RX} from adequately suppressing the non-orthogonal interference, causing the sensing performance to diverge from the theoretical bounds. A similar trend is observed in the communication \ac{NMSE} and \ac{BER} results depicted in Fig.~\ref{fig:mkf_metrics}(b) and Fig.~\ref{fig:mkf_metrics}(c). While $M=16$ results in notably poorer recovery overall, the spectral nulling variants demonstrate slightly higher resilience than their non-nulling counterparts. By explicitly puncturing sensing pilots to safeguard communication pilots, these architectures provide structural shielding for the \ac{CTX} channel estimation. 

\begin{figure}
\centering
\subfigure[]
{\includegraphics[width=0.47\textwidth]{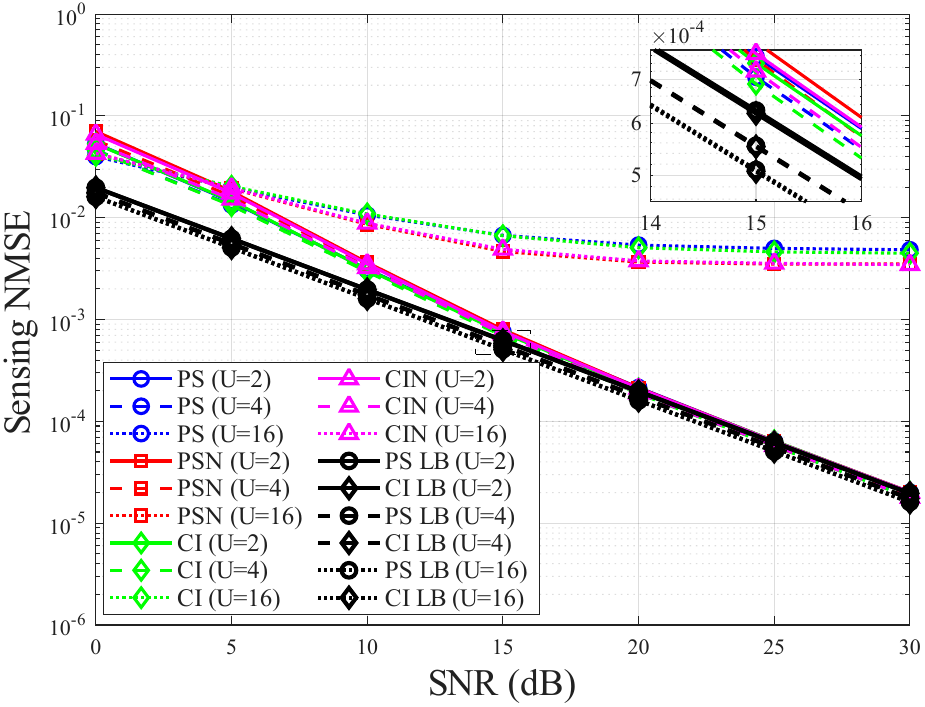}}
\subfigure[]
{\includegraphics[width=0.47\textwidth]{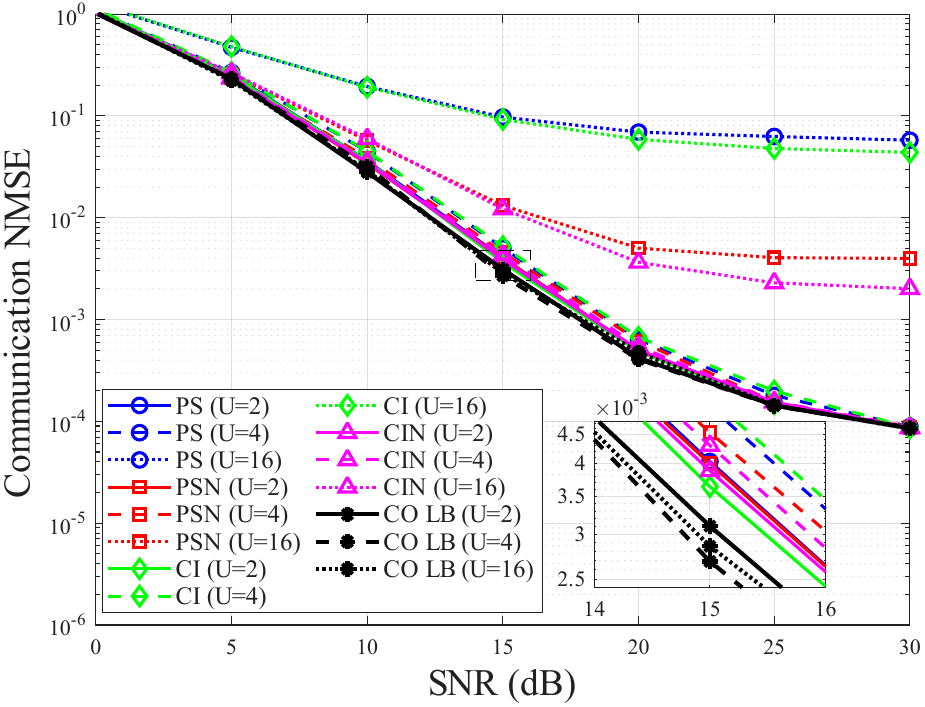}}
\subfigure[]
{\includegraphics[width=0.47\textwidth]{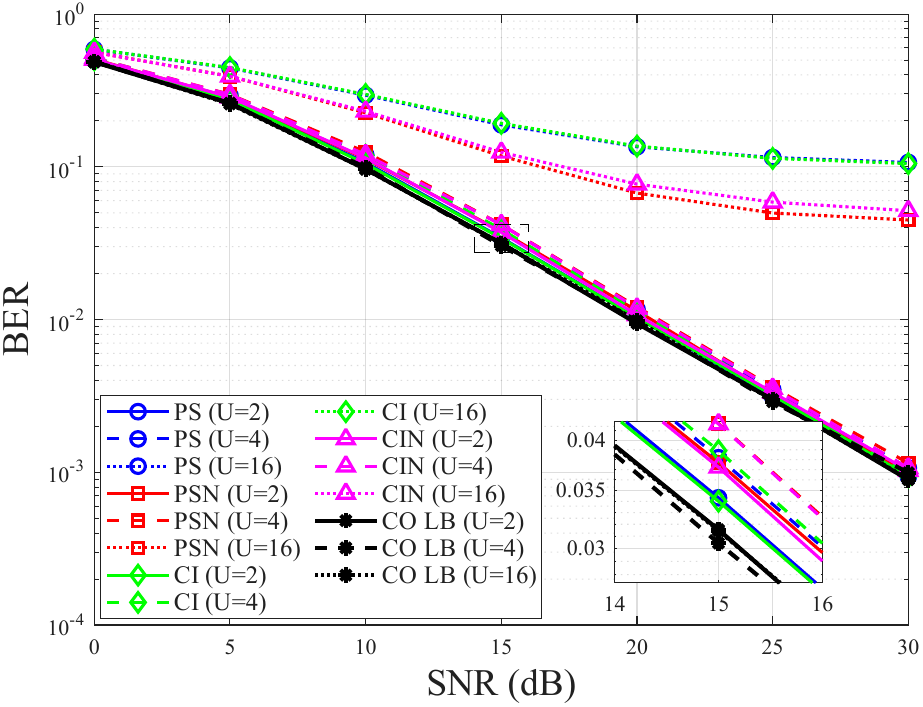}}
\caption{Scalability analysis of the \ac{IIC} architectures for $U \in \{2, 4, 16\}$ and $M=4$: explicitly subfigure (a) sensing \ac{NMSE}, subfigure (b) communication \ac{NMSE}, and subfigure (c) \ac{BER}. The lower bounds CO LB, PS LB, and CI LB represent the theoretical performance limits for the communication-only scenario as well as the PS-ISAC and CI-ISAC baseline configurations.}
\label{fig:mmf_metrics}
\end{figure}

\begin{figure}
\centering
\includegraphics[width=0.47\textwidth]{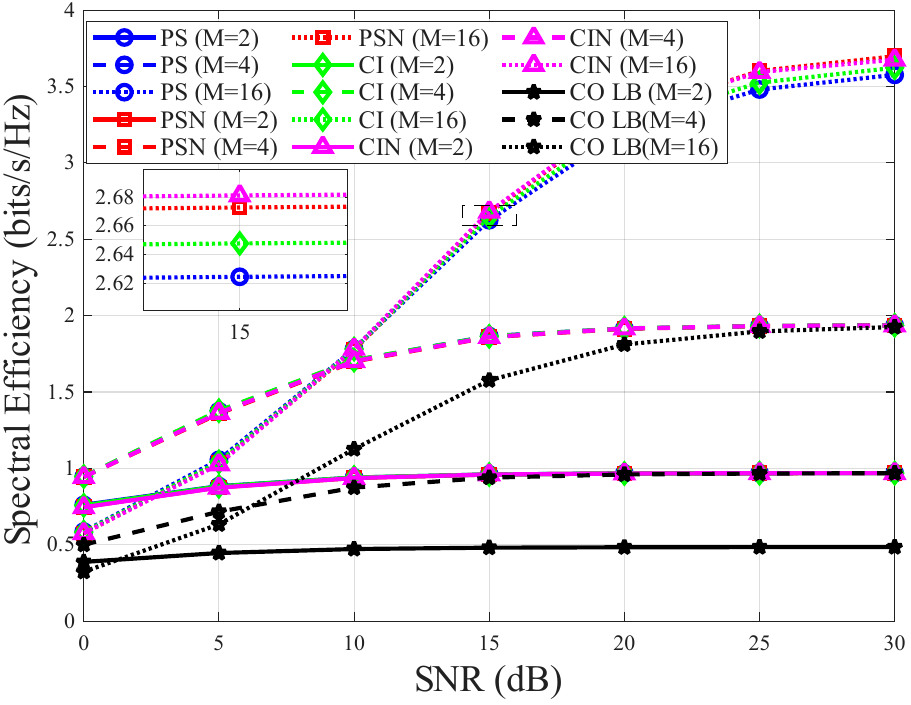}
\caption{\ac{SE} analysis sweeping the \ac{CTX} modulation order $M \in \{2, 4, 16\}$ for a fixed density of $U=4$ scheduled \acp{STX}. The evaluated \ac{NOMA} architectures provide a distinct throughput advantage over the orthogonal \ac{TDM} benchmark, explicitly denoted as the CO LB.}
\label{fig:se_m_sweep}
\end{figure}

\begin{figure}
\centering
\includegraphics[width=0.47\textwidth]{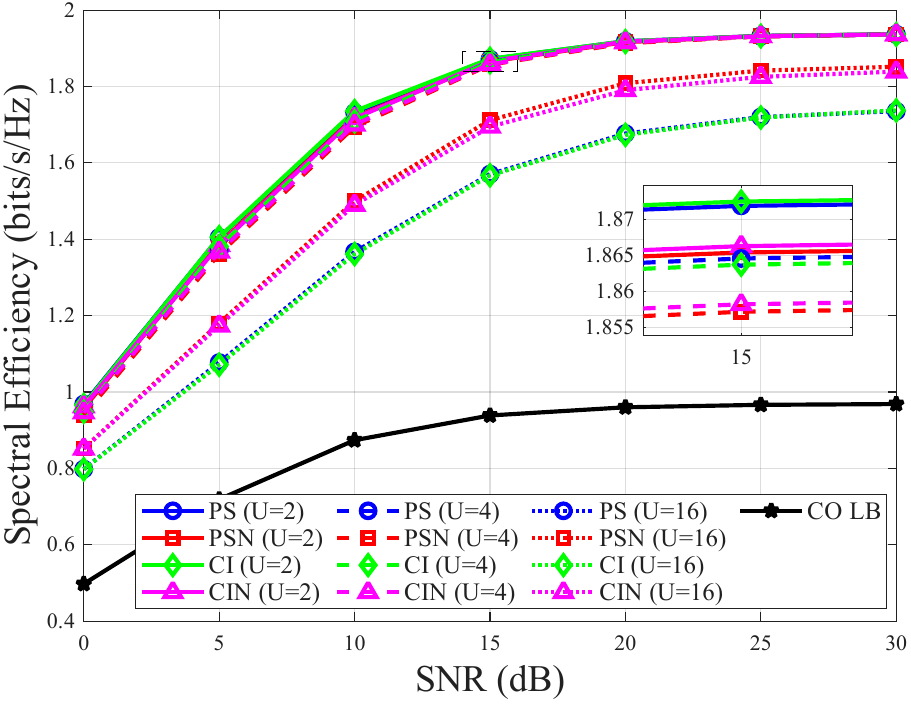}
\caption{\ac{SE} analysis under dense multi-STX scaling, specifically evaluating network densities of $U=2$, $U=4$, and $U=16$ scheduled \acp{STX} under a fixed modulation order of $M=4$.}
\label{fig:se_k_sweep}
\end{figure}

Furthermore, scalability in dense \ac{IoT} scenarios is rigorously evaluated by sweeping the number of scheduled \acp{STX} across $U \in \{2, 4, 16\}$ under a fixed modulation order of $M=4$. As shown in Fig.~\ref{fig:mmf_metrics}(a), scaling the network to an extreme density of $U=16$ fundamentally saturates the interference cancellation capabilities, leading to a severe performance departure from the sensing LBs. Nevertheless, for moderate densities ($U \leq 4$), the proposed \ac{NOMA} architectures approximate the theoretical sensing limits very closely.
The impact of device density on payload recovery is further confirmed by the communication \ac{NMSE} and \ac{BER} evaluations in Fig.~\ref{fig:mmf_metrics}(b) and Fig.~\ref{fig:mmf_metrics}(c). At $U=16$, the non-nulling overlapping structures fail to resolve the communication channel entirely due to the multi-STX interference. Conversely, the spectral nulling variants maintain operational stability and facilitate robust payload recovery under high-density interference scenarios.

\subsection{Spectral Efficiency}
To evaluate capacity, \ac{SE} trends across varying modulation orders and \ac{STX} densities are analyzed in Fig.~\ref{fig:se_m_sweep} and Fig.~\ref{fig:se_k_sweep}. Results show that \ac{NOMA}-based architectures, specifically PS, PSN, CI, and CIN, provide a distinct throughput advantage over the orthogonal \ac{TDM} baseline CO-ISAC. While the CO-ISAC benchmark equally splits time-domain resources between sensing and communication, the \ac{NOMA} frameworks improve spectral efficiency by enabling \ac{CTX} and \acp{STX} to share the same time-frequency resources simultaneously.

The modulation sweep in Fig.~\ref{fig:se_m_sweep} illustrates the fundamental trade-off between raw throughput and data reliability. As indicated by the \ac{BER} performance in Fig.~\ref{fig:mkf_metrics}, the $M=16$ configuration suffers from significant degradation with an error floor near $10^{-1}$. Despite this elevated \ac{BER}, $M=16$ yields the highest aggregate \ac{SE} by encoding four bits per subcarrier. This suggests that high-order modulations maximize throughput within \ac{ISAC}-\ac{NOMA} frameworks, while lower-order modulations remain essential for reliability-oriented applications. Moreover, the $M=2$ and $M=4$ configurations closely approach their theoretical upper bounds of one and two bits per channel use. In contrast, the severe non-orthogonal interference experienced at $M=16$ prevents the system from reaching its theoretical limit of four bits per channel use.
 
\begin{table*}[t]
\centering
\caption{Computational Complexity for Default Simulation Parameters ($N=512$, $U=4$, $Q=7$, $Q_1=3$, $Q_2=4$)}
\label{tab:numeric_complexity}
\renewcommand{\arraystretch}{1.5}
\begin{tabular}{|l|l|c|c|c|c|}
\hline
\multirow{2}{*}{\textbf{Architecture}} & \multirow{2}{*}{\textbf{IIC Strategy}} & \multicolumn{2}{c|}{\textbf{STX Side}} & \multicolumn{2}{c|}{\textbf{ISAC RX Side}} \\ \cline{3-6} 
 & & \textbf{Real Additions} & \textbf{Real Multiplications} & \textbf{Real Additions} & \textbf{Real Multiplications} \\ \hline\hline
\textbf{PS-ISAC-NOMA} & Joint ($7$ iters.) & $53,264$ & $20,496$ & $702,660$ & $279,748$ \\ \hline
\textbf{CI-ISAC-NOMA} & Joint ($7$ iters.) & $49,168$ & $12,304$ & $960,792$ & $344,344$ \\ \hline
\textbf{PSN-ISAC-NOMA} & Sequential ($3+4$) & $53,264$ & $20,496$ & $616,620$ & $242,860$ \\ \hline
\textbf{CIN-ISAC-NOMA} & Sequential ($3+4$) & $49,168$ & $12,304$ & $874,752$ & $307,456$ \\ \hline
\end{tabular}
\end{table*}

This trend is further corroborated by the network scaling analysis depicted in Fig.~\ref{fig:se_k_sweep} under a fixed modulation order of $M=4$. For moderate densities of $U=2$ and $U=4$ scheduled \acp{STX}, the \ac{NOMA} frameworks successfully suppress the mutual interference, enabling the system to almost perfectly reach the theoretical upper bound of two bits per channel use at high \ac{SNR} regimes. Conversely, scaling the architecture to an extreme density of $U=16$ scheduled \acp{STX} introduces significant multi-STX interference to the \ac{CTX} payload, significantly degrading the overall performance. Consistent with the \ac{BER} degradation observed in Fig.~\ref{fig:mmf_metrics}(c), the spectral nulling variants, explicitly PSN and CIN, provide superior \ac{SE} retention under this massive interference by structurally shielding the initial \ac{CTX} channel estimate.

While the interleaved nulling baseline ensures robust communication \ac{SE}, spectral sparsity inherently limits the maximum unambiguous range~\cite{APS-ISAC}. The proposed PS and PSN frameworks circumvent this bottleneck by employing phase-shifted pilots to maintain the optimal sensing resolution and maximum unambiguous range established in orthogonal designs~\cite{PS-ISAC}. These architectures leverage specialized \ac{IIC} receivers, particularly the sequential \ac{IIC} architecture for the PSN variant, to mitigate interference from dense multi-\ac{STX} superposition. Consequently, phase-shifted \ac{NOMA} achieves communication \ac{SE} levels comparable to interleaved baselines while preserving the underlying integrity of the sensing waveform. These results establish the proposed frameworks as scalable and efficient strategies for massive IoT \ac{ISAC} deployments.

\subsection{Computational Complexity Evaluation}
\label{subsec:sim_complexity}

To empirically validate the mathematical derivations presented in Section \ref{sec:complexity}, the exact hardware burden of the evaluated \ac{IIC} \acp{RX} is formally quantified. The evaluation considers an \ac{OFDM} framework utilizing $N=512$ subcarriers and serving $U=4$ scheduled \acp{STX}. To ensure a fair and consistent comparison, the iteration parameters $Q_1 = 3$, $Q_2 = 4$, and $Q = 7$ are selected.

Table \ref{tab:numeric_complexity} details the aggregate number of real additions and real multiplications required per \ac{OFDM} symbol duration, explicitly accounting for both the unitary transforms and the element-wise linear overheads. At the \ac{STX}, the proposed PS and PSN frameworks incur a marginal linear overhead of $8,192$ real multiplications and $4,096$ real additions for the four scheduled \acp{STX}. Because these phase-shift operations scale at $\mathcal{O}(UN)$, they are mathematically negligible compared to the dominant $\mathcal{O}(UN \log_2 N)$ transform costs, rendering the design highly suitable for resource-constrained \ac{IoT} edge devices.

The architectural advantages become heavily pronounced at the ISAC \ac{RX}. The CI framework utilizing separate per-\ac{STX} \acp{IFFT} blocks across the seven iterations, culminating in an exact arithmetic footprint of $344,344$ real multiplications. By substituting the separate per-\ac{STX} \acp{IFFT} with a single composite extraction array, the proposed PS architecture drastically reduces the transform count to forty-nine. This structural optimization yields an $18.8\%$ reduction in the total \ac{RX}-side multiplication burden, dropping from $344,344$ to $279,748$ operations.

Furthermore, the integration of the sequential \ac{IIC} architecture delivers significant computational savings for the spectral nulling variants. By fixing the initial \ac{CTX} channel estimate during the first three sequential cancellation iterations, the \ac{RX} entirely bypasses the complex overhead of iterative \ac{DFT}-based channel refinement during the initial phase. Consequently, the proposed PSN architecture executes merely forty-three total transform blocks over the entire \ac{IIC} loop. When evaluated directly against its interleaved counterpart, the proposed PSN framework achieves a rigorous $21.0\%$ reduction in exact \ac{RX}-side multiplications, scaling down from $307,456$ to $242,860$ operations. As a result, as explicitly validated in the preceding performance evaluations, the phase-shifted architectures secure these massive computational reductions without inflicting any degradation upon the sensing \ac{NMSE}, communication \ac{NMSE}, or \ac{BER} metrics. These exact arithmetic results conclusively prove that the proposed phase-shifted \ac{NOMA} frameworks provide a hardware-friendly, highly scalable solution perfectly tailored for energy-constrained, high-density \ac{IoT} deployments.

\section{Conclusion}
\label{sec:conclusion}
This paper presented a comparative analysis of multi-\ac{STX} \ac{UL} \ac{ISAC}-\ac{NOMA} networks utilizing phase-shifted sensing pilots and non-orthogonal communication. Two distinct \ac{IIC} architectures were evaluated: a joint \ac{IIC} for full-band configurations and a sequential \ac{IIC} for spectral nulling variants. Simulation results demonstrate that the spectral nulling architectures, \ac{PSN} and \ac{CIN}, provide superior communication \ac{NMSE}, \ac{BER}, and \ac{SE} as \ac{STX} density increases. While the PS framework facilitates convergence to interference-free lower bounds, the \ac{PSN} variant ensures robust \ac{SE} retention under high-order modulation. By shifting multi-\ac{STX} separation to the time domain, the proposed phase-shifted frameworks significantly reduce hardware complexity compared to interleaved baselines. Specifically, the \ac{PSN} framework is established as the optimal architecture for computational efficiency, followed by the PS framework as the secondary most efficient solution. These findings empirically validate the scalability of phase-shifted \ac{NOMA} for dense \ac{IoT} deployments.

Future investigations will derive a mathematical interference analysis to quantify performance boundaries as functions of \ac{CTX} modulation order and \ac{STX} density. Additionally, adaptive power allocation and machine learning-based suppression will be explored to enhance \ac{SE} and sensing resolution in dynamic \ac{IoT} environments.

\vfill

\clearpage
\end{document}